\documentclass[sigconf]{acmart}
\usepackage{hyperref}
\usepackage{graphicx}

\usepackage{lscape}
\usepackage{xcolor}
\usepackage{lmodern}
\usepackage{eurosym}
\usepackage{url}
\usepackage{tabularx} 
\usepackage[caption = false]{subfig}
\usepackage{tabularx}
\usepackage{multirow}
\usepackage{hhline}
\usepackage{hyperxmp}
\usepackage{tikz}
\usetikzlibrary{shapes.geometric, arrows.meta, positioning, shadows.blur, calc}

\newcolumntype{L}{@{}l@{}} 

\copyrightyear{2026} 
\acmYear{2026} 
\setcopyright{rightsretained} 

\acmConference[CAIN2026]{5th International Conference on AI Engineering -- Software Engineering for AI}{2026}{Rio de Janeiro (Brazil)}

\acmDOI{10.1145/xxx.xxxx}
\acmISBN{978-1-4503-xxx/xxx}

\begin{document}

\title{The Expert Validation Framework (EVF): \\Enabling Domain Expert Control in AI Engineering}


\author{Lucas Gren}
\affiliation{%
  \institution{Chalmers $|$ University of Gothenburg and \\Getinge AB}
    \city{Gothenburg}
  \country{Sweden}}
  \email{lucas.gren@lucasgren.com}

\author{Felix Dobslaw}
\affiliation{%
  \institution{Mid Sweden University}
    \city{Östersund}
  \country{Sweden}}
\email{felix.dobslaw@miun.se}

\begin{abstract}
Generative AI (GenAI) systems promise to transform knowledge work by automating a range of tasks, yet their deployment in enterprise settings remains hindered by the lack of systematic quality assurance mechanisms. We present an Expert Validation Framework that places domain experts at the center of building software with GenAI components, enabling them to maintain authoritative control over system behavior through structured specification, testing, validation, and continuous monitoring processes. Our framework addresses the critical gap between AI capabilities and organizational trust by establishing a rigorous, expert-driven methodology for ensuring quality across diverse GenAI applications. Through a four-stage implementation process encompassing specification, system creation, validation, and production monitoring, the framework enables organizations to leverage GenAI capabilities while maintaining expert oversight and quality standards.
\end{abstract}



\maketitle

\section{Introduction}
The integration of Generative AI (GenAI) technologies into software systems has created unprecedented opportunities for automating knowledge work across organizations \cite{staron2024bringing}. As the adoption of Large Language Models (LLMs) matures, industrial workflows are shifting from simple text-based interactions to complex software solutions capable of addressing repetitive tasks. While these systems are increasingly augmenting \cite{sergeyuk2025using} or even replacing traditional human-to-human interactions \cite{chatbotBook}, the industry is realizing the limitations of general-purpose models. Comprehensive, ``do-it-all'' platforms remain stubbornly out of reach for specialized professional tasks \cite{bespokeLegal,liang2024encouragingdivergentthinkinglarge}. Consequently, the focus has shifted toward engineering task- or domain-specific systems. This trend is particularly evident in the rise of autonomous agents (i.e., entities capable of taking action independently based on what it encounters in an ecosystem \cite{Wang_2024}), which offer significant potential for organizational needs \cite{xi2023risepotentiallargelanguage} but introduce new layers of non-deterministic behavior that require rigorous oversight.

While tools like GitHub Copilot have demonstrated impacts on developer productivity (AI for SE) \cite{chen2021evaluating,peng2023impact}, a broader challenge lies in Software Engineering for AI (SE4AI): building domain-specific applications where GenAI must perform expert-level tasks in fields like finance, law, or HR. This evolution builds on the classic idea of AI-powered question-answering systems \cite{lewis2020retrieval} but targets non-technical users. The latest such advance a the time of writing this paper is the release of Claude Cowork, which builds on Anthropic's successful coding agent, Claude Code, but provides a user interface designed specifically for business workflows rather than technical development.

Despite these advancements, a fundamental challenge persists: how can organizations ensure that AI-generated outputs maintain the accuracy, reliability, and domain-specific quality required for these mission-critical applications? As mentioned, this challenge becomes particularly pronounced as systems move toward increasing autonomy \cite{oviedo2024iso}. The trust gap between AI capabilities and organizational requirements reflects a misalignment between how AI systems demonstrate competence and how organizations assess quality \cite{brynjolfsson2018artificial}.

While GenAI systems excel at producing syntactically correct code or retrieving relevant information, determining whether those outputs are appropriate for specific organizational contexts remains fundamentally a domain expert judgment. Our work addresses this gap through a systematic framework that positions domain experts as primary stakeholders in AI quality assurance, transforming them from passive validators to active architects of system behavior.

\section{The Industrial Challenge}
While the potential of GenAI is vast, its application in regulated industrial settings reveals significant hurdles that are often invisible during the prototyping phase. In this section, we outline the primary obstacles organizations face when moving from proof-of-concept to production, specifically focusing on the gap between technical metrics and expert expectations.

\subsection{The Implementation Reality Gap}

In our experience, organizations deploying GenAI systems face a stark disconnect between impressive demonstrations and production-ready solutions. Consider a financial services firm implementing a RAG system (where the model retrieves specific proprietary data to ground its answers \cite{lewis2020retrieval}) for regulatory compliance queries. The system might retrieve relevant regulatory documents with high precision and generate coherent responses that appear technically sound. Yet, domain experts consistently identify critical issues: oversimplification of nuanced requirements, failure to acknowledge regulatory ambiguities that require human judgment, omission of critical caveats that practitioners would naturally include, and inappropriate tone for regulatory communications.

These deficiencies reflect a fundamental limitation: current GenAI systems lack the contextual understanding that domain experts develop through years of practice. Of course, working on what the LLM should use when solving a problem (i.e., context engineering --optimizing the prompt context with specific constraints and examples-- as opposed to only prompt engineering) could solve parts of this limitation, but not the fact that a knowledge base and a profession develops across time.  Automated metrics such as LLM-as-judge fail to capture these subtle but critical quality dimensions. 

The challenge intensifies with system evolution. GenAI systems exhibit inherent non-determinism, i.e., the same query can produce different outputs across invocations. Model providers frequently update their offerings, potentially changing system behavior in subtle ways. Organizations accumulate new knowledge and requirements continuously. Without systematic expert oversight, these factors compound to create quality drift that can undermine organizational trust.

\subsection{The Expert Knowledge Integration Problem}

Domain expertise encompasses far more than factual knowledge, and have (at least in research) been seen as key to actually solving problems with software \cite{fischer2009metadesign}. It includes understanding of organizational culture and implicit conventions, recognition of edge cases and their appropriate handling, awareness of political and regulatory sensitivities, and judgment about information relevance and appropriateness. This tacit knowledge, accumulated through experience and professional practice, proves difficult to codify into traditional software specifications.

In software development contexts, this manifests as AI-generated code that compiles successfully but violates architectural principles, creates technical debt, or introduces subtle security vulnerabilities that automated testing might miss \cite{chen2022codet}. Requirements engineering presents similar challenges, where AI assistance must navigate the complex interplay between stakeholder needs and technical constraints \cite{nascimento2023comparing}.

Furthermore, the dynamic nature of organizational knowledge compounds the challenge. Policies evolve, best practices advance, and regulatory landscapes shift. A GenAI system trained or configured with yesterday's knowledge may produce outputs that are technically correct but organizationally obsolete. The challenge extends beyond keeping information current to ensuring that system behavior evolves in alignment with expert understanding of how changes affect practice.

\section{The Expert Validation Framework}

\begin{figure}[htbp]
\centering
\resizebox{\linewidth}{!}{
\begin{tikzpicture}[
    node distance=1.2cm,
    process/.style={
        rectangle, 
        rounded corners, 
        minimum width=4cm, 
        minimum height=1.2cm, 
        text centered, 
        draw=black!70, 
        fill=aliceblue, 
        blur shadow={shadow blur steps=5},
        font=\sffamily\small
    },
    expert/.style={
        rectangle, 
        minimum width=3.5cm, 
        minimum height=0.8cm, 
        text centered, 
        draw=none, 
        fill=none, 
        font=\sffamily\bfseries\footnotesize,
        text=darkmidnightblue
    },
    arrow/.style={thick, ->, >=stealth, color=darkgray},
    line/.style={thick, color=darkgray},
    label/.style={font=\sffamily\tiny\itshape, text=darkgray, align=center, fill=white, inner sep=2pt}
]

\definecolor{aliceblue}{rgb}{0.94, 0.97, 1.0}
\definecolor{darkmidnightblue}{rgb}{0.0, 0.2, 0.4}
\definecolor{darkgray}{rgb}{0.3, 0.3, 0.3}


\node (spec) [process, align=center] {
    \textbf{1. Specification}\\
    \scriptsize Define Scope \& Quality Boundaries\\
    \scriptsize \textit{(Expert Ownership)}
};

\node (knowledge) [process, below=of spec, align=center] {
    \textbf{2. Knowledge Foundation}\\
    \scriptsize Curation, RAG Setup \& Context Eng.\\
    \scriptsize \textit{(Living Asset)}
};

\node (validation) [process, below=of knowledge, align=center] {
    \textbf{3. Socratic Validation}\\
    \scriptsize Expert-AI Dialogue \& Edge Case Discovery\\
    \scriptsize \textit{(3--5 Iterative Cycles)}
};

\node (monitor) [process, below=of validation, align=center] {
    \textbf{4. Production Monitoring}\\
    \scriptsize Drift Detection \& User Feedback\\
    \scriptsize \textit{(Continuous Dialogue)}
};


\draw [arrow] (spec) -- (knowledge);
\draw [arrow] (knowledge) -- (validation);
\draw [arrow] (validation) -- node[right, font=\tiny] {Approved} (monitor);

\draw [arrow] (validation.east) -- ++(0.8,0) |- node[label, near start] {Refinement Loop} (knowledge.east);

\draw [arrow] (monitor.west) -- ++(-0.8,0) |- node[label, near start, rotate=90] {New Edge Cases / Policy Changes} (spec.west);

\end{tikzpicture}
}
\caption{The Expert Validation Framework (EVF). The process relies on two critical feedback loops: the \textit{Socratic Refinement Loop} during development and the \textit{Continuous Adaptation Loop} from production back to specification.}
\label{fig:evf_process}
\end{figure}
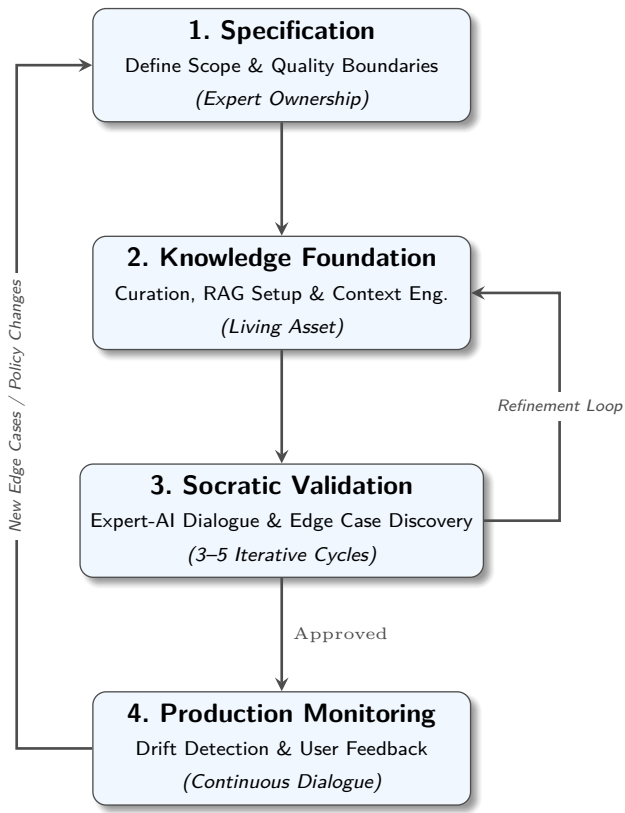

Our framework transforms the traditional AI deployment model by positioning domain experts as primary architects of system behavior rather than post-hoc validators. This shift reflects a fundamental insight: trust in GenAI systems cannot be achieved through technical metrics alone but requires systematic alignment between AI capabilities and expert knowledge throughout the development lifecycle. There are, of course, software testing systems that implement GenAI testing capabilities (like Google's Vertex AI\footnote{\url{https://docs.cloud.google.com/vertex-ai/generative-ai/}}), but these systems are primarily designed to provide enterprise-grade tools for the objective and data-driven assessment of the models themselves. While they successfully support technical development tasks (such as model migrations, prompt editing, and fine-tuning), they remain focused on optimization rather than domain correctness. Crucially, they do not place the domain expert in the driver's seat, leaving a gap in validating whether the system's behavior aligns with professional standards and organizational context. Our new framework's overall process flow, detailing the cyclical nature of these expert-driven interactions and the feedback loops between stages, is depicted in Figure \ref{fig:evf_process}.

\subsection{Core Design Principles}

The framework operationalizes three interconnected principles that govern all aspects of system development and deployment:

\textbf{Expert Authority as Foundation}. Domain experts maintain ultimate authority over quality standards within their domains. This principle extends beyond simple approval rights to encompass active participation in system design, test creation, and continuous monitoring. Technical teams provide implementation support but cannot override expert judgment on matters of domain correctness.

\textbf{Systematic Validation Through Structured Processes}. Quality assurance follows reproducible methodologies with clearly defined stages, criteria, and decision points. Each stage builds upon previous validations, creating a cumulative quality guarantee. This systematic approach enables organizations to audit quality decisions, identify improvement opportunities, and scale successful practices across domains.

\textbf{Continuous Monitoring and Adaptation}. System quality is not static but requires ongoing vigilance. Regular expert reviews, user feedback integration, and proactive knowledge maintenance ensure that systems remain aligned with evolving organizational needs. This principle acknowledges that even well-validated systems can experience quality drift through changing contexts or usage patterns.

\subsection{The Socratic Method in Practice}

A distinguishing feature of our framework is the application of Socratic dialogue to expert-AI collaboration. Rather than positioning AI as a solution provider that experts validate, we establish an iterative conversation where AI proposes approaches and actively solicits expert refinement. This ``two-way handshake'' ensures that expert knowledge shapes system behavior throughout development rather than serving merely as post-hoc validation.

Consider test case generation for a healthcare documentation system. The AI might propose: ``Based on the knowledge base, I've identified these potential test scenarios for patient intake documentation. Should we also test for cases where patients have multiple insurance providers? What about scenarios involving minors with divorced parents?'' This guided discovery helps experts articulate tacit knowledge while ensuring comprehensive coverage. The expert might also want to add additional considerations the AI did not identify.

This collaborative approach proves particularly valuable when dealing with edge cases and boundary conditions. By engaging in dialogue about potential failure modes, experts and AI systems jointly explore the problem space more thoroughly than either could achieve independently.

\section{Implementation Stages and Practical Insights}
To address the challenges of quality and trust, the EVF is structured around four distinct stages that guide the system from initial concept to long-term maintenance. Each stage is designed to maximize the capture of tacit expert knowledge while maintaining technical rigor.

\paragraph{Stage 1: Specification Through Expert Lens}

Implementation begins with rigorous specification that captures not just functional requirements but also the contextual nuances that define quality in specific domains. In our experience so far, organizations often underestimate the importance of this foundational stage, rushing toward technical implementation without adequately defining success criteria.

As an example, consider the legal domain where the specification process could reveal distinctions between ``legally correct'' and ``organizationally appropriate'' responses. Domain experts in HR, as another example,  might say that certain technically accurate HR interpretations should never be provided without explicit disclaimers about seeking counsel with an HR expert, even though the AI system might generate technically correct summaries. These boundaries, invisible to technical teams, are essential for responsible deployment.

The specification stage also establishes clear ownership models. Each knowledge domain receives assigned experts who maintain ongoing accountability for quality. This assignment goes beyond nominal responsibility, i.e., these experts actively participate in all subsequent stages, ensuring continuity of domain understanding throughout development.

\paragraph{Stage 2: Knowledge Foundation as Living Asset}

The knowledge foundation represents more than a static document repository. Our experience so far shows that successful systems treat knowledge curation as an ongoing investment requiring active expert maintenance. In general, the initial knowledge audit often triggers comprehensive documentation updates across an organization. This is a challenge since there is more work by the domain experts until  they have to do less work. Getting a knowledge base in good shape takes time. 

Technical representation are also surprisingly important. Converting poorly formatted documents into structured formats (such as Markdown or XML) can dramatically improve system performance. In one HR deployment, transforming complex Word documents with embedded tables and inconsistent formatting into clean Markdown improved answer accuracy substantially.

The collaborative nature of knowledge foundation development surfaces implicit organizational knowledge. Experts frequently discover undocumented practices or informal conventions that significantly impact system quality. These discoveries, while initially adding complexity, ultimately strengthen both the AI system and organizational knowledge management practices.

\paragraph{Stage 3: Validation Beyond Accuracy}

Test case development shows the multifaceted nature of quality in enterprise AI systems. Our framework emphasizes that validation must assess not just factual correctness but also appropriateness, completeness, tone, and adherence to organizational conventions. This comprehensive approach to validation distinguishes production-ready systems from technical demonstrations.

The iterative refinement process typically requires 3--5 cycles before systems meet expert standards. Each iteration addresses progressively subtle issues. Initial cycles often identify fundamental problems like missing information sources, incorrect retrieval priorities, or inappropriate response generation. Later cycles refine nuances such as tone, qualification statements, and edge case handling.

\paragraph{Stage 4: Production Monitoring as Continuous Dialogue}

Deployment marks a transition from development to operational excellence. Our framework treats production monitoring as an active process requiring sustained expert involvement. Regular spot-checks, user feedback review, and periodic comprehensive assessments ensure that systems maintain quality over time. If the domain experts are in control of the systems and its quality, as suggested, it is not a matter of validating the system with experts, instead it must be part of their job to maintain the AI system. 

Real-world usage consistently shows patterns not anticipated during testing. In an HR system deployment, employees might frequently ask questions about policy interactions, e.g., how vacation policies interacted with parental leave, for example that might not be covered in initial test cases. These discoveries drive continuous test suite expansion, ensuring that validation coverage evolves with actual usage patterns.

The feedback loop between production usage and test suite enhancement are particularly valuable. Each identified quality issue becomes a new test case, gradually building comprehensive coverage of real-world scenarios. This approach transforms user feedback from complaint management into systematic quality improvement.

\section{Lessons from Early Implementations}
Deploying the EVF in real-world settings has provided valuable insights into the intersection of human expertise and AI behavior. Beyond the technical implementation, we have identified several organizational and human-centric factors that critically influence the success of GenAI adoption.

\subsection{The Expert Capability Challenge}

A significant challenge emerges in finding domain experts with sufficient technical knowledge to effectively participate in the framework. Traditional domain experts may lack familiarity with test suite management, version control, or systematic validation processes. Conversely, technically skilled staff often lack the deep domain knowledge necessary to assess output quality.

We think this needs a major change in who we hire to different departments. We believe that some domain experts, like HR, Finance, etc, needs to be tech-savvy enough to build and maintain the test suites for whatever AI is intended to solve problems for them in their domain of expertise.  

The development of no-code tools for test management and validation could reduce technical barriers. Visual interfaces for test case creation, drag-and-drop knowledge organization, and simplified approval workflows can enable broader expert participation. Investment in these accessibility improvements by technical teams pays dividends through increased expert ownership and more comprehensive quality coverage. We think the software developers should focus on building a platform that the domain experts can use without having to write or edit any code. 

A common implementation hurdle is disagreement between experts. Unlike rule-based systems, GenAI outputs exist on a spectrum of validity. We found that the validation phase often surfaces conflicting interpretations of organizational policy among senior experts. If there is no clear single owner of the knowledge base, which there often is though, the framework forces these conflicts to be resolved in the knowledge foundation phase, effectively turning the AI engineering process into a consensus-building mechanism for the organizational knowledge.

\subsection{Trust Through Transparency}

For an organizations to successfully build trust in GenAI systems we think they must have a radical transparency about validation processes. Rather than treating AI quality as a black box, these organizations actively communicate what experts tested, how quality standards were defined, what monitoring occurs, and how feedback drives improvement. The LLMs are not auditable in the same way a classical IT systems due to their complexity and must be tested differently.

The transparency extends to acknowledging system limitations. When experts clearly communicate boundaries, i.e., what the system can and cannot do, users hopefully develop appropriate expectations and trust. We think that if the domain experts show how they tested the system and what it is suppose to handle, the other experts within the same domain will trust the system more initially. Of course, the system must prove itself across time, which is also what we are trying to address.  The framework itself becomes a trust-building tool. Knowing that domain experts maintain ongoing oversight and can intervene if quality degrades provides organizational confidence in AI deployment.

\section{Implications for AI Engineering Practice}

Our framework challenges conventional approaches to AI system development by repositioning quality assurance from a technical concern to a domain expertise challenge. This shift has profound implications for how organizations structure AI initiatives, allocate resources, and measure success.

The framework suggests that sustainable AI deployment requires ongoing investment in expert participation rather than one-time technical implementation. Organizations must not only budget for, but target their recruitment to include domain experts with technical skills for continuous expert involvement and knowledge foundation maintenance. These ongoing costs, often overlooked in initial AI planning, are essential for maintaining system quality over time.

Furthermore, the framework highlights the importance of treating AI systems as living entities requiring active management rather than static deployments. Just as software systems require maintenance, updates, and enhancement, AI systems demand continuous expert attention to maintain alignment with organizational needs. What we suggest is that software engineering for AI struggles without the domain experts driving the maintenance of the AI system, and the software engineers focus on enabling the domain experts to build AI systems instead of building AI systems for them. 

\section{Conclusion and Future Directions}
The Expert Validation Framework suggests a structured approach to bridging the gap between GenAI capabilities and organizational quality requirements. By positioning domain experts as primary stakeholders in AI quality assurance, the framework enables organizations to leverage AI capabilities while maintaining the oversight necessary for trustworthy deployment.

We have conducted early tests, however, this framework needs further implementation and assessment, which we are working on. 
We now focus on developing no-code tools that further reduce barriers to expert participation, automated test case generation, and learning what processes are translatable across non-technical domains. As AI capabilities continue to evolve, we believe that frameworks that systematically incorporate human expertise become increasingly critical for responsible and effective deployment.

The path forward requires recognizing that AI quality in enterprise contexts is fundamentally a human judgment challenge, at least for now. Technical metrics and automated testing, while necessary, are insufficient for ensuring that AI systems meet the complex, context-dependent requirements of real-world deployment. By placing experts at the center of AI quality assurance, organizations can realize GenAI's transformative potential while maintaining the standards necessary for mission-critical applications. We think a broad deployment of general-purpose chatbots are good as brainstorming partners for staff, but for an expert to delegate a task to an AI, we need significantly more sophisticated ways to test system behavior to assess if the quality is good enough for that specific use case.

\bibliographystyle{ACM-Reference-Format}
\bibliography{references}

\end{document}